\documentstyle[12pt,epsfig]{article}
\voffset     = - 0.10 in
\begin{document}
\thispagestyle{empty}


{\large\bf
\hfill
\begin{tabular}{r@{}}
\end{tabular}
}

\vspace*{1cm}

{\Large\bf
\begin{center}
Analytical Representation 
of the Longitudinal Hadronic Shower Development
\end{center}
}

\bigskip


\bigskip

\begin{center}
{\large\bf Y.A.~Kulchitsky} 

\smallskip
Institute of Physics, National Academy of Sciences, Minsk, Belarus
\\
\&
JINR, Dubna, Russia 

\bigskip

{\large\bf V.B.~Vinogradov}

\smallskip
JINR, Dubna, Russia
\end{center}

\vspace*{\fill}

\begin{abstract}
The analytical representation of the longitudinal hadronic sho\-wer development
from the face of a calorimeter is presented and compared with 
experimental data.
The suggested formula is particularly useful at designing, 
testing and calibration of huge calorimeter complex like in ATLAS at LHC.  
\vskip 5mm 
\noindent
\end{abstract}

\vspace*{\fill}

\newpage

One of the important questions of hadron calorimetry is the question
of the longitudinal development of hadronic showers
\cite{bock97}.
For analysing of energy depositions in various depths of calorimeters
it is necessary to have the analytical representation of longitudinal shower 
development from the face of calorimeter.   
Such formula is particularly useful at designing, testing and calibration 
of huge calorimeter complex like in ATLAS \cite{atcol94} at LHC.  

There is well known parameterisation of longitudinal shower development 
from shower origin suggested by Bock et al.\ \cite{bock81}:
\begin{equation}
        \frac{dE_{s} (x)}{dx} = 
                k\      
                \Biggl\{
                w\ \biggl( \frac{x}{X_o} \biggr)^{a-1}\ 
                e^{- b \frac{x}{X_o}}\
                + \ 
                (1-w)\ 
                \biggl( \frac{x}{\lambda_I} \biggr)^{a-1}\ 
                e^{- d \frac{x}{\lambda_I}}
                \Biggr\}, 
\label{elong00}
\end{equation}
where 
$X_o$ is the radiation length, 
$\lambda_I$ is the interaction length,
$a,\ b,\ d,\ w$ are parameters, 
$k$ is the normalisation factor. 
In this work we use the values of parameters obtained in 
\cite{hughes90}: 
$a = 0.786 + 0.427 \cdot lnE$,
$b = 0.29$,
$d = 0.978$,
$\omega = 1.032 - 0.159 \cdot lnE$.

In practice, the calorimeter longitudinal segmentation is unsufficient 
for precise determination of shower vertex position.
Because of it is necessary to integrate the longitudinal profile from shower
origin over the shower position:
\begin{equation}
        \frac{dE (x)}{dx} = 
        \int \limits_{0}^{x} 
        \frac{dE_{s}(x-x_v)}{dx}\ e^{- \frac{x_v}{\lambda_I}}\ dx_v ,
\label{elong01}
\end{equation}
where $x_v$ is a coordinate of the shower vertex.

In given work the analytical representation of the hadronic shower 
longitudinal development from the calorimeter face has been derived.

It turns out that 
\begin{eqnarray}
        \frac{dE (x)}{dx} & = & 
                N\ 
                \Biggl\{
                \frac{w X_o}{a} 
                \biggl( \frac{x}{X_o} \biggr)^a 
                e^{- b \frac{x}{X_o}}
                {}_1F_1 \biggl(1,a+1, 
                \biggl(b - \frac{X_o}{\lambda_I} \biggr) \frac{x}{X_o}
                \biggr) 
                \nonumber \\
                & & + \ 
                \frac{(1 - w) \lambda_I}{a} 
                \biggl( \frac{x}{\lambda_I} \biggr)^a 
                e^{- d \frac{x}{\lambda_I}}
                {}_1F_1 \biggl(1,a+1,
                - \bigl( 1 - d \bigr) \frac{x}{\lambda_I} \biggr)
                \Biggr\} , 
\label{elong03}
\end{eqnarray}
where 
${}_1F_1(\alpha,\beta,z)$ 
is the confluent hypergeometric function \cite{abramovitz64} and
$N$ is the normalisation factor.
The confluent hypergeometric function in the first term
of formula (\ref{elong03}) can be calculated by using the following
relation:
\begin{equation}
        {}_1F_1 (1,a+1,z) =
                a z^{-a} e^{z}\ \gamma (a,z),
\label{elong05}
\end{equation}
where $\gamma (a,z)$ is the incomplete gamma function \cite{abramovitz64}. 
This function can be calculated by using the entry G106 in the program library 
\cite{cernlib}.
The confluent hypergeometric function in the second term
can be calculated by using the series development:
\begin{equation}
        {}_1F_1 (1,a+1,-z) =
                1 - \frac{z}{a+1} + \frac{z^2}{(a+1)(a+2)} + \dots .
\label{elong06}
\end{equation}
This series has the remarkable property: at $a = 0$ it corresponds to the 
function $e^{-z}$.
For real calorimeters with longitudinal size usually less then 10 $\lambda_I$
the value of $z = (1-d) \frac{x}{\lambda_I}$ does not exceed $0.2$.   
Therefore it is sufficient only 3 terms in order to calculate the 
function  ${}_1F_1 (1,a+1, -z)$ with the precision of 0.1\%. 

The normalisation factor is  
\begin{equation}
        N = E_{beam} \biggr/ \int \limits_{0}^{\infty} \frac{dE(x)}{dx} dx ,
\label{elong040}
\end{equation}
where $E_{beam}$ is the beam energy.
The integral in the denominator can be calculated by using the following
relation \cite{gradshteyn}:
\begin{equation}
        \int \limits_0^{\infty} 
        z^{\nu} 
        e^{-\lambda z} 
        {}_1F_1 (\alpha, \gamma, k z)\ dz =
        \Gamma (\nu +1) {\lambda}^{- \nu -1} 
        {}_2F_1 \biggl(\alpha ,\nu +1, \gamma , \frac{k}{\lambda} \biggr),   
\label{elong07}
\end{equation}
where ${}_2F_1(\alpha,\beta,\gamma, z)$ 
is the hypergeometric function \cite{abramovitz64}, 
which in our case have the very simple form:
\begin{equation}
        {}_2F_1 (1,a+1,a+1,z) = \frac{1}{1-z} .
\label{elong08}
\end{equation}
As a result we obtain
\begin{equation}
        N =     
                \frac{E_{beam}}{\lambda_I\ \Gamma (a)\ 
                \bigl(w\ X_o\ b^{-a} + (1-w)\ \lambda_I\ d^{-a} \bigr)
                } . 
\label{elong04}
\end{equation}

\begin{figure*}[tbph]
     \begin{center}
        \begin{tabular}{c}
\mbox{\epsfig{figure=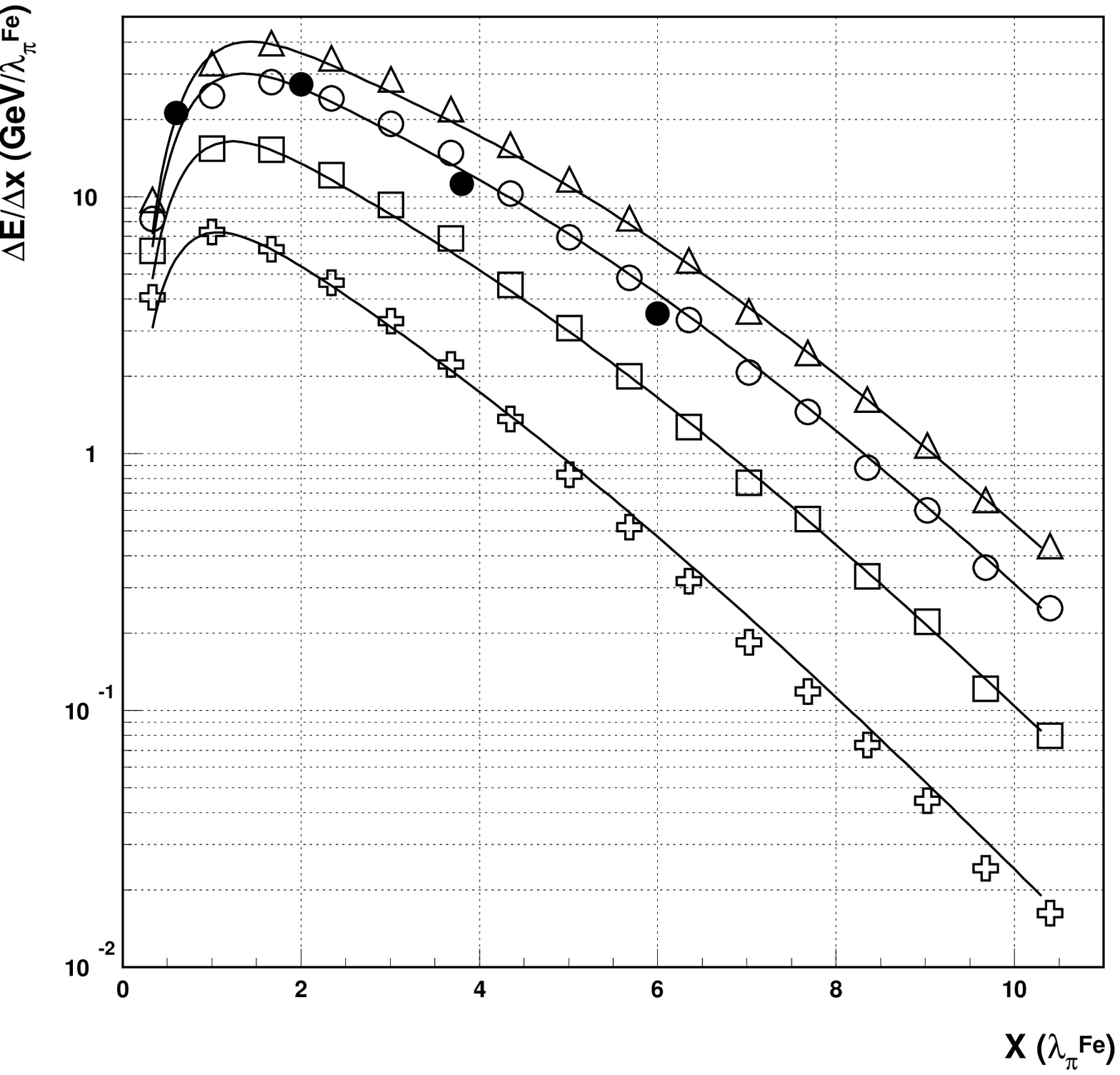,width=0.95\textwidth,height=0.75\textheight}} 
        \\
        \end{tabular}
     \end{center}
       \caption{
        Longitudinal profiles of the hadron showers of 20 GeV (crosses),
        50 GeV (squares), 100 GeV (open circles) and 140 GeV (triangles)
        energies as a function of the longitudinal coordinate $x$ in units 
        $\lambda_I$ for conventional iron-scintillator 
        calorimeter \cite{hughes90} and of 100 GeV (black circles)
        for tile iron-scintillator calorimeter \cite{budagov97}. 
        The solid lines are calculations by function (\ref{elong03}) 
        with parameters from \cite{hughes90}.
       \label{f1}}
\end{figure*}

In Fig.\ \ref{f1} the calculations by formula (\ref{elong03}) are shown 
and compared with the experimental data at 20 -- 140 GeV obtained by
using the conventional iron-scintillator calorimeter \cite{hughes90} and 
the one with innovative feature --- the longitudinal orientation of the 
scintillating tiles \cite{budagov97}.
The good agreement is observed.

So, now there is useful analytical formula which gives the possibility 
immediately to obtain the longitudinal shower energy deposition 
from calorimeter face. 

\newpage


\newpage


\end{document}